\documentclass[twocolumn,aps,showpacs]{revtex4}
\usepackage{amsmath}
\usepackage{amssymb}
\usepackage{epsf}

\begin{document}

\title{Selective transmission of Dirac electrons and ballistic
magnetoresistance of \textit{n-p} junctions in graphene.}
\author{Vadim V. Cheianov and Vladimir I. Fal'ko}
\affiliation{Department of Physics, Lancaster University, Lancaster, LA1 4YB, United
Kingdom}

\begin{abstract}
We show that an electrostatically created \textit{n-p} junction separating
the electron and hole gas regions in a graphene monolayer transmits only
those quasiparticles that approach it almost perpendicularly to the \textit{%
n-p} interface. Such a selective transmission of carriers by a single 
\textit{n-p} junction would manifest itself in non-local magnetoresistance
effect in arrays of such junctions and determines the unusual Fano factor in
the current noise universal for the \textit{n-p} junctions in graphene.
\end{abstract}

\maketitle

The chiral nature of quasiparticles in graphene monolayers and bilayers \cite%
{Haldane,DiVincenzo,Semenoff,DresselhausBook,AndoNoBS,McCann} has been
revealed in several recent experiments \cite%
{novo04,novo05,novo05pnas,zhang05,NovoselovMcCann}. The Fermi level in a
neutral graphene sheet (a monolayer of carbon atoms with hexagonal lattice
structure) is pinned near the corners of the hexagonal Brillouin zone \cite%
{kpoints} which determine two non-equivalent valleys in the quasiparticle
spectrum. \ The quasiparticles in each of the two valleys are described by
the Hamiltonian \cite{DresselhausBook,Semenoff}, 
\begin{equation*}
\hat{H}_{1}=v\,\mathbf{\sigma }\cdot \mathbf{p},
\end{equation*}%
where the 'isospin' Pauli matrices $\sigma _{i}$ operate in the space of the
electron amplitude on two sites (A and B) in the unit cell of a hexagonal
crystal \cite{kpoints}, $\mathbf{p}=(p_{x},p_{y})=-i\mathbf{\nabla }$ is the
momentum operator \cite{hbar1} defined with respect to the centre of the
corresponding valley, and $v$ is a constant formed by the A-B hopping \cite%
{DresselhausBook}. The Dirac-type Hamiltonian $\hat{H}_{1}$ determines the
linear dispersion $vp$ for the electron, and $-vp$ for the hole branch of
quasiparticles. In each valley \cite{kpoints}, the electron and hole states
also differ by the isospin projection onto the direction of their momentum:
electrons have chirality $\mathbf{\sigma }\cdot \mathbf{p}/p=1$, holes $%
\mathbf{\sigma }\cdot \mathbf{p}/p=-1$. Therefore, in structures where the
quasiparticle isospin is conserved (, a monolayer with electrostatic
potential scattering) their backscattering is strictly forbidden \cite%
{AndoNoBS}, which gives rise to the peculiar properties of the \textit{n-p}
junction in graphene reported in this Letter.

Since an atomically-thin graphitic film is a gapless semiconductor, carrier
density in it can be varied using external gates \cite{novo04} from
electrons to holes \cite{novo04,novo05,novo05pnas,zhang05,NovoselovMcCann}.
A planar \textit{n-p} junction in graphene can be made, \textit{e.g.}, using
split-gates, and in view of a rapidly improving mobility of the new material 
\cite{novo05,novo05pnas,zhang05} it may soon be possible to fabricate
ballistic circuits of electrically controlled graphene-based \textit{n-p}
junctions. Below, we model the \textit{n-p} junction in graphene using the
electrostatic potential $u(x)=vk_{F}\eta (x/d)$ characterized by a single
length scale $d$ and the Fermi momentum $k_{F}$ determined by the equal
densities of the electron and hole gases on the opposite sides of it. Here $%
\eta (\pm \infty )=\pm 1$, $\eta ^{\prime }(0)=1$, and the line $x=0$
separates the \textit{p-} and \textit{n-}regions. Since in a junction
produced by electrostatic gates the length $d$ \ is about the inter-gate
distance and exceeds the electron wavelength in a monolayer, we focus this
study on smooth \textit{n-p} junctions with $k_{F}d>1$, and show that their
transmission properties are determined by the central region where $%
u(x)\approx Fx$ [$F=vk_{F}/d$].

\begin{figure}[t]
\centerline{\epsfxsize=1.0\hsize \epsffile{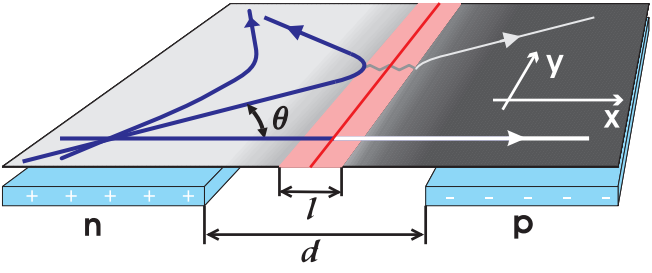}}
\caption{Angular dependence of quasiparticle transmission through the
electrostatically generated \textit{n-p} junction in graphene. }
\label{Fig1}
\end{figure}

The transport properties of the \textit{n-p} junction are determined by the
angular dependence of the probability $w(\theta )$ of an electron incident
from the left with an energy equal to the chemical potential $\mu =0$ to
emerge as a hole on the right hand side of the junction. The electron
approaching the center of the junction has the kinetic energy $v\sqrt{%
p_{x}^{2}+p_{y}^{2}}$, where $p_{y}=k_{F}\sin \theta $ is conserved. Due to
the energy conservation, the $x$-component of the electron momentum is $%
p_{x}(x)=\sqrt{u^{2}(x)/v^{2}-p_{y}^{2}}$, the classically allowed region
for the electron motion is determined by the condition $|u|>p_{y}v$ and its
trajectory cannot extend beyond the turning point at the distance $%
l=vp_{y}/F $ from the center of the junction \cite{footnote-momentum}. For a
particle incident perpendicular to the junction ($p_{y}=0$) the classically
forbidden region disappears. Moreover, due to the isospin conservation which
prohibits backscattering of chiral quasiparticles \cite{AndoNoBS}, the wave
incident at $\theta =0$ is perfectly transmitted, though, for any small $%
\theta $, the transmission probability is determined by tunnelling through
the classically forbidden region, $w\sim e^{-2S}$, where $%
S=i\int_{-l}^{l}p_{x}(x)dx=\frac{1}{2}\pi vp_{y}^{2}/F$. For a smooth 
\textit{n-p} junction shown in Fig. \ref{Fig1} with $F=vk_{F}/d$ and $%
k_{F}d\gg 1$, this yields (for $\theta \ll \pi /2$) 
\begin{equation}
w(\theta )=e^{-\pi (k_{F}d)\sin ^{2}\theta }.  \label{w}
\end{equation}

The angular dependence of the transmission probability given in Eq. (\ref{w}%
) is, in fact, exact in the range $\theta \ll \pi /2$ 
for any smooth junction and represents the central
result of this Letter. Below, we rigorously derive the result in Eq. (\ref{w}%
) using the method of transfer matrix and also show that $w(\theta )=\cos
^{2}\theta $ for a step-like potential. Similarly to the formulae \cite%
{Glazman} describing adiabatic ballistic constrictions in semiconductors,
the applicability of Eq. (\ref{w}) is not restricted by the constraint $w\ll
1$. This can be used to describe how a smooth \textit{n-p} junction
selectively transmits only carriers approaching it within a small angle $%
\theta \lesssim \theta _{0}=1/\sqrt{\pi k_{F}d}$ around the perpendicular
direction and to determine the conductance per unit length of a broad
junction, 
\begin{equation}
g_{np}=\frac{4e^{2}}{h}\int \frac{k_{F}d\theta }{2\pi }w(\theta )\approx 
\frac{2e^{2}}{\pi h}\sqrt{\frac{k_{F}}{d}},  \label{G}
\end{equation}%
and the universal Fano factor \cite{footnoteBeen} in the shot noise, 
\begin{equation}
\langle \langle I\cdot I\rangle \rangle =(1-\sqrt{\tfrac{1}{2}})\,eI.
\label{Fano}
\end{equation}%
At the end of this paper we shall discuss several ballistic
magnetoresistance effects which exploit the selectivity of transmission
implicit in Eq. (\ref{w}).

To formulate the scattering problem, we shall exploit the separation of $x$
and $y$ variables for the electron motion across the junction (in $x$
direction) and the fact that momentum along $y$-axis (parallel to the
junction) is conserved. This makes the scattering problem one-dimensional
(1D). The scattering states at the energy equal to the chemical potential, $%
\mu =0$ are spinors satisfying the Dirac-type equation 
\begin{equation}
-i\partial _{x}\sigma _{x}\psi +v^{-1}u(x)\psi +p_{y}\sigma _{y}\psi =0,
\label{dirac}
\end{equation}%
which conserves the 1D current $J_{x}=\psi ^{\dagger }\sigma _{x}\psi $.

To find the transmission probability $w(\theta )$ for such states, we
calculate the transfer matrix $T(x,y)$ which satisfies the equation 
\begin{equation}
\partial _{x}T(x,y)=L(x)\,T(x,y),\;L=-i\frac{u(x)}{v}\sigma _{x}+p_{y}\sigma
_{z}  \label{TL}
\end{equation}%
and the conditions $T(y,y)=I,$ $T(x,y)=T(x,z)T(z,y),$ $\det T(x,y)=1,$ and $%
T^{\dagger }(x,y)\sigma _{x}T(x,y)=\sigma _{x}$. To relate the transmission
coefficient $w$ to the transfer matrix $T(x,y)$ one has to factor out the
asymptotic evolution of the reflected and transmitted waves. This can be
done by using matrices $A_{\pm }$ satisfying the wave equation in the
asymptotic regions, 
\begin{equation}
\partial _{x}A_{\pm }(x)=(\mp ik_{F}\sigma _{x}+p_{y}\sigma _{z})A_{\pm }(x),
\label{Adef}
\end{equation}%
such that their columns are made of right- and left-propagating states
normalized to carry unit current. The explicit expression for these matrices
is 
\begin{equation*}
A_{\pm }(x)=\sqrt{\frac{k_{F}}{2p_{x}}}\left( 
\begin{array}{ll}
\frac{p_{x}\pm ip_{y}}{k_{F}}e^{\mp ip_{x}x} & \frac{-p_{x}\pm ip_{y}}{k_{F}}%
e^{\pm ip_{x}x} \\ 
e^{\mp ip_{x}x} & e^{\pm ip_{x}x}%
\end{array}%
\right) ,
\end{equation*}%
where $p_{x}(x)=\sqrt{u^{2}(x)/v^{2}-p_{y}^{2}}=\sqrt{k_{F}^{2}-p_{y}^{2}}.$
Then, the transmission probability can be found using the matrix 
\begin{equation}
\left( 
\begin{array}{cc}
\alpha & \beta ^{\ast } \\ 
\beta & \alpha ^{\ast }%
\end{array}%
\right) \equiv \lim_{x\rightarrow \infty
}A_{+}^{-1}(x)T(x,-x)A_{-}(-x),\;\;w=\frac{1}{|\alpha |^{2}}.  \label{Mon}
\end{equation}

To illustrate the transfer matrix formalism, we calculate the probability of
a Dirac fermion transmission through a sharp potential step $u(x)=vk_{F}%
\mathrm{sign}(x).$ In this case, we factor the transfer matrix as $%
T(x,y)=T_{+}(x,0)T_{-}(0,y)$, where $T_{+(-)}(x,y)$ is a transfer matrix on
the right(left) side of the junction, each given by $T_{\pm }(x,y)=A_{\pm
}(x)A_{\pm }^{-1}(y)$. Using this solution and Eq. (\ref{Mon}) we find 
\begin{equation}
\left( 
\begin{array}{cc}
\alpha & \beta ^{\ast } \\ 
\beta & \alpha ^{\ast }%
\end{array}%
\right) =A_{+}^{-1}(0)A_{-}(0),\;\alpha =1-\frac{ip_{y}}{\sqrt{%
k_{F}^{2}-p_{y}^{2}}}.  \notag
\end{equation}%
For the transmission probability this yields 
\begin{equation}
w=1-(p_{y}/k_{F})^{2}=\cos ^{2}\theta ,
\end{equation}%
which manifests the chiral nature of quasiparticles in graphene. Indeed, the
free electron states of the Dirac Hamiltonian $\hat{H}_{1}$ have their
isospin polarized along the momentum (for the transmitted holes, with $%
\mathbf{p}=(-k_{F}\cos \theta ,k_{F}\sin \theta )$, it is antiparallel), and
the reflection amplitude of an electron is determined by the scalar product $%
\psi _{\nearrow }^{\dagger }\cdot \psi _{\nwarrow }\sim \sin \theta $ of its
initial and final state spinors.

To calculate the transmission probability for a smooth potential with $%
k_{F}d\gg 1$, we separate the $x$-axis across the junction into the inner (%
\textit{i}) and outer (\textit{o}) parts. In the outer part, $\left\vert
x\right\vert >cd$ (where $c\ll 1$), we find the T-matrix, $T_{o}$ using the
method of adiabatic expansion. Then, we match it with the exact solution, $%
T_{i}$ obtained in the central part of the junction, $\left\vert
x\right\vert <d$, where the potential $u(x)$ can be linearised, $u(x)\approx 
{k_{F}x}/{d}$, and obtain the complete tranfer matrix as $%
T(y,x)=T_{o}(x,a)T_{i}(a,-a)T_{o}(-a,y)$.

For the adiabatic expansion of the transfer matrix $T_{o}$ we use a
transformation 
\begin{equation}
Y(x)=\frac{1}{u}\left( 
\begin{array}{ll}
i\varkappa & i\varkappa ^{\ast } \\ 
u & u%
\end{array}%
\right) ,\;\left\{ 
\begin{array}{c}
p_{x}(x)=\sqrt{\dfrac{u^{2}}{v^{2}}-p_{y}^{2}}, \\ 
\varkappa =p_{y}+ip_{x}(x).%
\end{array}%
\right.  \label{Ydef}
\end{equation}%
which locally diagonalises the $L$ operator in Eq. (\ref{TL}): 
\begin{equation}
Y^{-1}LY=ip_{x}(x)\sigma _{z}.  \label{qdef}
\end{equation}%
The transfer matrix $\tilde{T}_{o}$ defined in a new basis, 
\begin{equation}
T_{o}(x,y)=Y(x)\tilde{T}_{o}(x,y)Y^{-1}(y),  \label{tildeTdef}
\end{equation}%
satisfies the equation 
\begin{gather}
\partial _{x}\tilde{T}_{o}(x,y)=ip_{x}(x)\sigma _{z}\tilde{T}%
_{o}(x,y)+\Omega (x)\tilde{T}_{o}(x,y),  \label{tildeT} \\
Q=-Y^{-1}\partial _{x}Y=\frac{p_{y}u^{\prime }(x)}{2p_{x}^{2}(x)u(x)}\left( 
\begin{array}{ll}
-\varkappa & \varkappa ^{\ast } \\ 
\varkappa & -\varkappa ^{\ast }%
\end{array}%
\right) .  \notag
\end{gather}%
In the adiabatic approximation the matrix $\Omega (x)$ is assumed to be
small as compared to the diagonal term $p_{x}(x)\sigma _{z}$, and to the
leading order Eq.~(\ref{tildeT}) is solved by 
\begin{equation}
\tilde{T}_{o}(x,y)=\exp \left[ i\sigma _{z}\int_{y}^{x}p_{x}(x^{\prime
})dx^{\prime }\right] .  \label{Tildetans}
\end{equation}

Formally, the adiabatic approximation is justified if $\left\vert
p_{y}u^{\prime }/(up_{x}^{2})\right\vert \ll 1$, which breaks down near the
turning points $p_{x}(x)=0$ and when $u(x)=0$. However, for the junctions
with $k_{F}d\gg 1$, the interval between turning points lies within the
region of space where the the potential profile can be approximated using
the linear function $u(x)={k_{F}x}/{d}$. The transfer matrix in this region, 
$T_{i}$ can be found from Eq. (\ref{TL}) exactly, using the transformation 
\begin{gather}
T_{i}(x,y)=e^{-i\frac{\pi }{4}\sigma _{y}}e^{-i\frac{\phi (x)}{2}\sigma _{z}}%
\tilde{T}_{i}(x,y)e^{i\frac{\phi (y)}{2}\sigma _{z}}e^{i\frac{\pi }{4}\sigma
_{y}},  \label{T1def} \\
\mathrm{where}\;\;\phi (x)=k_{F}d^{-1}x^{2}.  \notag
\end{gather}%
This is because the matrix $\tilde{T}_{i}$ satisfies the equation 
\begin{equation}
\partial _{x}\tilde{T}_{i}(x,y)=-p_{y}\left( 
\begin{array}{ll}
0 & e^{i\phi (x)} \\ 
e^{-i\phi (x)} & 0%
\end{array}%
\right) \tilde{T}_{i}(x,y),  \label{T1eq}
\end{equation}%
where the upper row of $\tilde{T}_{i}$ can be expressed in terms of two
linearly independent solutions of the equation 
\begin{equation*}
e^{i\phi }\partial _{x}e^{-i\phi }\partial _{x}\Psi =p_{y}^{2}\Psi ,
\end{equation*}%
while the lower row can be expressed in terms of their complex conjugate.
Equation (\ref{T1eq}) is symmetric with respect to the parity transformation 
$x\rightarrow -x$, and its even/odd solutions are 
\begin{align*}
\Psi _{\text{even}}(x)& =\Phi \left( {\textstyle-i\frac{p_{y}^{2}d}{4k_{F}}},%
\textstyle{\ \frac{1}{2}};\,i\phi \right) , \\
\Psi _{\text{odd}}(x)& =-p_{y}x\Phi \left( {\textstyle\frac{1}{2}-i\frac{%
p_{y}^{2}d}{4k_{F}}},\textstyle{\ \frac{3}{2}};\,i\phi \right) .
\end{align*}%
where $\Phi $ is the confluent hypergeometric (Kummer) function \cite%
{AbrSteg} with the following asymptotic properties:%
\begin{equation*}
\Phi (a,b;z\rightarrow i\infty )\approx \frac{\Gamma (b)}{\Gamma (b-a)}\frac{%
e^{i\pi a}}{z^{a}}+\frac{1}{\Gamma (a)}e^{z}z^{a-b}.
\end{equation*}%
Therefore, inside the interval $|x|,|y|<cd$ the transfer matrix $\tilde{T}%
_{i}$ can be written as 
\begin{equation}
\tilde{T}_{i}(x,y)=B(x)B^{-1}(y),\;\;B=\left( 
\begin{array}{cc}
\Psi _{\text{even}} & \Psi _{\text{odd}} \\ 
\Psi _{\text{odd}}^{\ast } & \Psi _{\text{even}}^{\ast }%
\end{array}%
\right) ,  \label{T1np}
\end{equation}%
where the matrix $B$ satisfies Eq. (\ref{T1eq}) and has unit Wronskian, $%
\text{det}\,B=1$.

Finally, after a chain of substitutions, the obtained solutions for the
matching transfer matrices $T_{o}$ and $T_{i}$ can be combined together into 
\begin{equation*}
T(y,x)=T_{o}(x,a)T_{i}(a,-a)T_{o}(-a,y),
\end{equation*}%
and used to calculate the parameters $\alpha $ and $\beta $ in Eq. (\ref{Mon}%
), 
\begin{align*}
\alpha & =e^{\frac{\pi p_{y}^{2}d}{2k_{F}}}, \\
\beta ^{\ast }& =-e^{\frac{\pi p_{y}^{2}d}{4k_{F}}}\frac{\sqrt{2\pi }e^{i%
\frac{\pi }{4}}\left( \frac{p_{y}^{2}d}{2k_{F}}\right) ^{\frac{1}{2}+\frac{%
ip_{y}^{2}d}{2k_{F}}}}{\Gamma (1+\frac{ip_{y}^{2}d}{2k_{F}})}e^{i\chi }, \\
\chi & =p_{x}(\infty )l-\int_{l}^{\infty }[p_{x}(x^{\prime })-p_{x}(\infty
)]dx^{\prime },
\end{align*}%
needed for determining the transmission probability, 
\begin{equation}
w=|\alpha |^{-2}=e^{-\pi p_{y}^{2}d/k_{F}}.  \label{w-final}
\end{equation}

A selective transmission of carriers by a smooth \textit{n-p} junction
described by Eqs. (\ref{w-final},\ref{w}), with $k_{F}d\gg 1$, only allows
for the passage of quasiparticles approaching the junction in an almost
perpendicular direction, with $p_{y}<\sqrt{k_{F}/d}\ll k_{F}$ and $\theta
\ll 1$. This makes the transport characteristics of ballistic graphene-based
devices sensitive to the geometrical orientation of \textit{n-p} junctions
in them, and it is capable of generating a sizeable magnetoresistance (MR)
effect.

\begin{figure}[t]
\centerline{\epsfxsize=1.0\hsize \epsffile{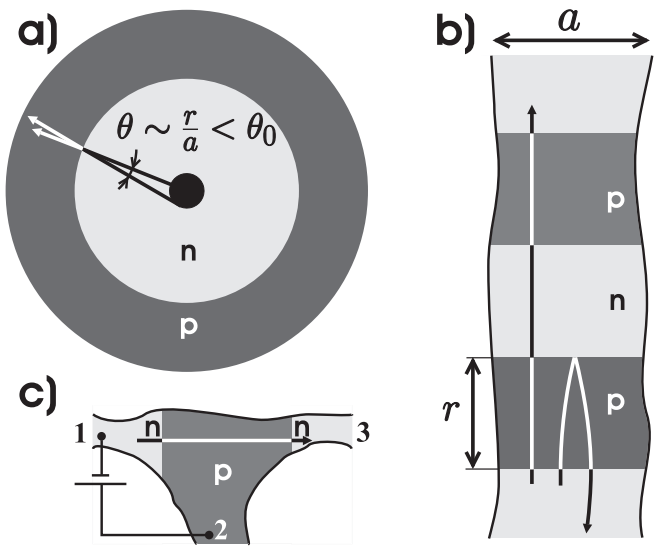}}
\caption{Ballistic MR devices with \textit{n-p} junctions in graphene: (a)
Corbino geometry; (b) series of \textit{n-p-n} junctions, with the
illustration of trajectories of electrons transmitted by the first junction
for $B=0$ (left) and $B>B\ast $ (right); (c) three-terminal cavity. }
\label{Fig2}
\end{figure}

A nominal resistance, $R_{\mathrm{np}}=1/ag_{np}$ of a single, separately
taken \textit{n-p} junction with the peripheral length $a$ separating the
electron and hole gases with densities $n_{e/h}=k_{F}^{2}/\pi $ is
determined by Eq. (\ref{G}). Whether or not the nominal junction resistance
contributes to the total resistance of a ballistic device depends on how
free carriers propagate in it. For example, when an \textit{n-p} junction,
with the perimetre $a=2\pi r$, separates two metallic Corbino contacts to
the ballistic 2D electron/hole gases shown in Fig. \ref{Fig2}(a), electrons
emitted from the inner contact with the radius $b<r/\sqrt{\pi k_{F}d}$ reach
the junction at the incidence angle $\theta <\theta _{0}=1/\sqrt{\pi k_{F}d}$
and pass it without scattering. As a result, the presence of the \textit{n-p}
junction does not affect the Corbino resistance, unless an external magnetic
field changes the incidence angle to $\theta ^{\prime }=r/r_{c}\gtrsim
\theta _{0}$, where $r_{c}=k_{F}\hbar c/eB$\ is the cyclotron radius in the
ballistic region. This generates the MR, 
\begin{gather}
R(B)=R_{\text{ext}}+\frac{f(B/B_{\ast })}{ag_{np}},\;\mathrm{where}%
\;f(0)=0,\;f(1)\sim 1,  \notag \\
\mathrm{and}\;\;\;B_{\ast }=\left( \hbar c/e\right) \sqrt{k_{F}/\pi r^{2}d}.
\label{MR1}
\end{gather}

A strong MR effect can also be expected an a Hall-bar sample with several
parallel \textit{n-p-n} junctions, Fig. \ref{Fig2}(b). The energy-averaged 
\cite{footnoteEnAv} transmission through the series of two junctions, $%
w_{2}(\theta )=\left[ w^{-1}(\theta )+w^{-1}(\theta +r/r_{c})-1\right] ^{-1}$
is determined by the individual junction transmissions $w(\theta )$ and $%
w(\theta +\frac{r}{r_{c}})$. Here, we take into account that, due to the
external magnetic field, an electron transmitted by the first junction at
the incidence angle $\theta $ would approach the second at the angle $\theta
^{\prime }=\theta +\frac{r}{r_{c}}$, where $r/r_{c}=B/B_{\ast }$ with $%
B_{\ast }$ defined in Eq. (\ref{MR1}). In the absence of a field $\theta
^{\prime }=\theta $, and the transmitted particle would also pass the second
junction, as shown on the left of Fig. \ref{Fig2}(b). If, due to a high
field, the angle $\theta ^{\prime }$ is sufficient for the particle to be
reflected, $\theta ^{\prime }>\theta _{0}$ the latter would return to the
first junction along the path illustrated on the right in Fig. \ref{Fig2}(b)
and escape to the contact where it came from. This would suppress the
conductance of the \textit{n-p-n} junction down to the value determined
scattering by the side edges of the sample. \ Having substituted $%
w_{2}(\theta )$ [instead of $w(\theta )$] into the conductance per unit
length of a broad junction defined in Eq. (\ref{G}), we find the
magnetoconductance of the \textit{n-p-n} junction, 
\begin{equation}
g_{npn}(B)\approx \frac{g_{np}}{\sqrt{\pi }}\int_{-\infty }^{\infty }\frac{dx%
}{e^{x^{2}}+e^{(x+\frac{B}{B_{\ast }})^{2}}-1}.  \label{MR2}
\end{equation}

A strongly selective quasiparticle transmission in Eqs. (\ref{w},\ref%
{w-final}) can also be used for creating ballistic cavity-type structures in
graphene, with non-local transport properties. In a three-terminal 'cavity'
shown in Fig. \ref{Fig2}(c), a \textit{p}-charging gate would produce two
parallel \textit{n-p} junctions, so that ballistic electrons emitted from
the contact 1 and transmitted by the first junction would easily pass
through the second and reach contact 3. As a result, a bias voltage applied
between contacts 1 and 2 would generate current between contacts 1 and 3,
thus giving rise to the trans-conductance $G_{12}^{13}$ with a strong
magnetic field dependence, 
\begin{equation}
G_{12}^{13}(B)\sim \frac{2e^{2}}{\pi h}\sqrt{\frac{a^{2}k_{F}}{d}}\;f(\tfrac{%
B}{B_{\ast }}).  \label{trans-cond}
\end{equation}

The authors thank T. Ando, A. Geim, J. Jefferson, and E. McCann for useful
discussions. \ This project has been supported by the Lancaster-EPSRC
Portfolio Partnership EP/C511743.


\begin{thebibliography}{99}
\bibitem{kpoints} Corners of the hexagonal Brilloin zone are $\mathbf{K}%
_{\xi }=\xi ({\textstyle\frac{4}{3}}\pi a^{-1},0)$, where $\xi =\pm 1$ and $%
a $ is the lattice constant. \ To use the same Dirac-type Hamiltonian in
both valleys, we determine the two-component iso-spinor as $\psi _{\xi
=+1}=(\phi (A),\phi (B))$ in the valley $K$ ($\xi =+1$) and $\psi _{\xi
=-1}=(\phi (B),\phi (A))$ in the valley $\tilde{K}$ ($\xi =-1$). \ In such
notations, time reversal is described by $(\Pi _{1}\otimes \sigma _{x}){\hat{%
H}}^{\ast }\left( {\mathbf{p}}\right) (\Pi _{1}\otimes \sigma _{x})={\hat{H}}%
\left( -\mathbf{p}\right) $, and spatial inversion (parity) by $\Pi _{1}{%
\hat{H}}\left( {\mathbf{p}}\right) \Pi _{1}={\hat{H}}\left( -{\mathbf{p}}%
\right) $, where $\Pi _{1}$ swaps valleys.
\bibitem{hbar1} Everywhere below we shall use units $\hbar =1$.
\bibitem{footnote-momentum} Note that the transmitted quasiparticle on the
other side of the junction have momentum $p_{x}=-\sqrt{%
u^{2}(x)/v^{2}-p_{y}^{2}}$.
\end{thebibliography}
\end{document}